\documentclass[journal]{IEEEtran}

\ifCLASSINFOpdf
  
\else
  
\fi

\usepackage[svgnames]{xcolor}
\definecolor{BurntOrange}{rgb}{0.8,0.33,0.0}
\usepackage{cite}

\usepackage{standalone}

\usepackage{amsmath,amsfonts,amssymb,mathtools}
\usepackage{algorithm}
\usepackage{tikz}
\usepackage{booktabs}
\usetikzlibrary{decorations.pathreplacing}
\usetikzlibrary{arrows.meta}
\definecolor{cIn}  {RGB}{167,195,221}   % readings, light
\definecolor{cInD} {RGB}{ 30, 86,143}   % readings, deep
\definecolor{cOut} {RGB}{238,199,149}   % voltage targets
\definecolor{cOut2}{RGB}{234,184,118}   % angle-difference targets
\definecolor{cOutD}{RGB}{176, 98, 15}   % targets, deep
\definecolor{cQ}   {RGB}{203,178,218}   % unknown, light
\definecolor{cQD}  {RGB}{104, 56,140}   % unknown, deep
\definecolor{cNet} {RGB}{170,208,178}   % network, light
\definecolor{cNetD}{RGB}{ 27,106, 60}   % network, deep
\definecolor{sxH}{RGB}{ 58, 71, 80}    % header, slate
\definecolor{sxA}{RGB}{ 15,110,120}    % Test A, teal
\definecolor{sxB}{RGB}{168, 37, 71}    % Test B, raspberry
\definecolor{sxD}{RGB}{154,107,  8}  
\usetikzlibrary{shadows}

\usepackage[
  colorlinks=true,          % flip to hidelinks for camera-ready (see toggle below)
  linkcolor=blue,
  citecolor=blue,
  urlcolor=blue,
  breaklinks=true,          % allow links to break across lines/columns
  bookmarks=false,          % IEEE PDF eXpress prefers no bookmark tree
  pdfpagemode=UseNone,
  hypertexnames=false,      % required: IEEEtran reuses equation counters
  pdfstartview={FitH},
  pdfauthor={Ayushi Jolotia and Parikshit Pareek},
  pdftitle={State Estimation as In-context Tabular Prediction},
  pdfsubject={IEEE Transactions on Power Systems},
  pdfkeywords={power system state estimation, foundation model, in-context learning, weighted least squares}
]{hyperref}
\usepackage[scaled=0.99,zerostyle=c]{newtxtt}
\usepackage[capitalise,noabbrev]{cleveref}   % MUST come after hyperref
\crefname{theorem}{Theorem}{Theorems}
\crefname{corollary}{Corollary}{Corollaries}
\crefname{assumption}{Assumption}{Assumptions}
\crefname{remark}{Remark}{Remarks}

\begin{document}

\title{State Estimation as In-context Tabular Prediction}

\author{Ayushi Jolotia~\IEEEmembership{Student Member,~IEEE}, Parikshit Pareek$^\dagger$,~\IEEEmembership{Member,~IEEE}
       % <-this % stops a space
\thanks{\noindent $^\dagger$Corresponding Author;$\,$ Authors are with the Department of Electrical Engineering, Indian Institute of Technology Roorkee. \textit{\{ayushi\_j;pareek\}@ee.iitr.ac.in}; The authors acknowledge the funding support provided by the ANRF PM Early Career Research Grant and the IIT Roorkee Faculty Initiation Grant. \vspace{-1em}}}

\maketitle

\begin{abstract}
This letter poses state estimation (SE) as an in-context tabular prediction (ICTP) problem. The key idea is a tabular arrangement in which historical measurement vectors and corresponding state estimates serve as the context, and a new measurement vector forms the query for which we predict a state vector. We propose to use tabular foundation models (TFMs), pretrained on synthetic data, whose weights are never updated, so power system data is needed only as context. With frozen TFM weights we estimate the state of 14-, 30-, 33-, 70-, and 118-bus transmission and distribution networks, and do so where weighted least squares (WLS) fails or degrades, under low observability and with bad data. ICTP errors are 11 to 57$\times$ below the per-bus average of the context at full metering, and 1.3 to 4.0$\times$ below it with only 5\% of measurements available.
With measurements at six buses biased by 8$\sigma$, the ICTP angle error rises 1.8 to 5.0$\times$ over its bias-free value against 2.5 to 12.7$\times$ for WLS with bad-data detection.
Without retraining, ICTP has lower voltage error than DSS$^2$ on both of its feeders, and lower than a retrained TAPINE on two of three reconfiguration sets.
\end{abstract}


\IEEEpeerreviewmaketitle
\vspace{-1.3em} 
\section{Introduction}
\label{sec:introduction}
The state of a power system, the voltage magnitude and phase angle at every bus, underpins monitoring, control, and planning. It is estimated from meter readings that carry measurement and communication errors. The standard estimator, weighted least squares (WLS), minimizes the weighted sum of squared residuals, which is the maximum-likelihood estimate when errors are zero-mean Gaussian with known covariance. WLS relies on two conditions that are not guaranteed in practice. First, errors must be Gaussian. A gross error from a stuck meter or a mis-scaled transducer is still weighted by its meter's assumed variance. Residual-based $\chi^2$ and largest-normalized-residual tests are applied afterward, but a gross error at a leverage point pulls the estimate onto itself and escapes detection \cite{abur2004}. Second, the system must be observable. Meters go out of service or lose communication, and WLS fails abruptly once observability is lost.

Recently, learned estimators have been proposed that map measurements to states using historical data. DSS$^2$ trains a graph neural network by penalizing power-flow violations, without seeing true states \cite{habib2024dss2}, while TAPINE builds the physics into its architecture \cite{tapine2026}. Both are tied to their training network: a new network, measurement configuration, or topology requires retraining.
The same is true for Bayesian and data-driven estimators, which infer the state of unobservable systems from historical data \cite{mestav2019, zamzam2019} and require network-specific model training.
This letter addresses both the WLS failure modes and the need for network-specific training by posing SE as an in-context tabular prediction (ICTP) problem. The main idea is that, given historical measurements with their estimated states, the current state can be inferred by a pretrained tabular foundation model (TFM). TFMs such as TabPFN v2~\cite{tabpfn} and TabICL v2.1~\cite{tabicl} are transformer-based in-context learners, pretrained on synthetic tables, learning how columns relate to one another and to the target. Given SE data in this tabular format as context, the model predicts the label of a new query in a single forward pass. 
We stress-test ICTP under measurement loss, noise variation, bad data, and topology change, against classical and learned estimators.
Our main contributions are:
\vspace{-0.2em} 
\begin{enumerate}
    \item \textbf{State estimation as in-context tabular prediction.} Measurements form the input columns, and per-bus voltage magnitudes and per-line angle differences form the target columns. 
    No step uses the measurement Jacobian or the admittance matrix. Each TFM uses a single pretrained checkpoint across all five networks.

    \item \textbf{Estimation without an observability precondition.} ICTP returns a full state vector at every measurement level tested. WLS can fail with more measurements than states. On the 118-bus system it returns no estimate once half the meters are gone, with 362 readings for 235 states. On the radial 33-bus feeder with a single unmetered bus, WLS returns angle estimates worse than each bus's historical average, even with bad-data checks. Over the same dark-bus sweep, the ICTP angle error grows at most 4.4$\times$ over its full-metering value on all three networks.

    \item \textbf{Robustness to bad data and evaluation on published benchmarks.} 
    With six buses biased by $8\sigma$, the WLS angle error rises 2.5 to 12.7$\times$ against 1.8 to 5$\times$ for ICTP. On TAPINE's reconfiguration sets, where TAPINE degrades by 19 and 35$\times$ in voltage once the feeder is reconfigured beyond what it was trained on, the TabPFN voltage error varies by at most 1.2$\times$ across all three sets.
\end{enumerate}
\vspace{-1.em} 
\section{Proposed In-Context State Estimation}
\label{sec:proposed-ice}
WLS minimizes the weighted squared residuals between the measurements and the nonlinear measurement function $\mathbf{h}(\mathbf{x})$, solved iteratively by Gauss-Newton using the gain matrix $\mathbf{G}=\mathbf{H}^{\top}\mathbf{R}^{-1}\mathbf{H}$, with $\mathbf{H}$ the Jacobian of $\mathbf{h}$ and $\mathbf{R}$ the noise covariance \cite{abur2004}. Each measurement contributes one row to $\mathbf{H}$, and observability is decided by the rank of $\mathbf{H}$, since $m \ge n$ is necessary but not sufficient. When $\mathbf{H}$ loses full column rank, $\mathbf{G}$ becomes singular, and the normal equations have no unique solution. With finite precision, the WLS solver may fail outright or return an inaccurate estimate.

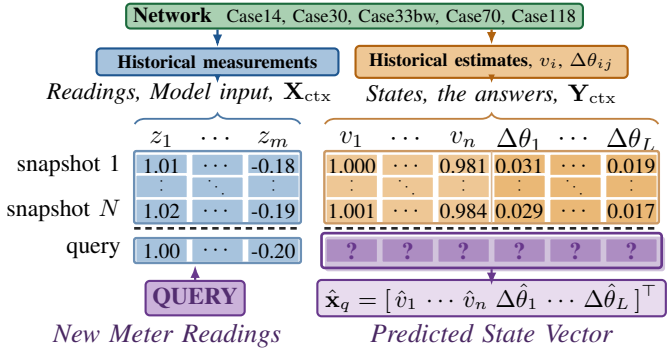
\begin{figure}[!t]
\centering
\vspace{-2.1em} 
\begin{tikzpicture}[
  x=7.4mm, y=3.5mm,
  cell/.style={draw=white, line width=1.0pt, minimum width=7.0mm, minimum height=3.3mm,
               inner sep=0pt, font=\footnotesize},
  zc/.style ={cell, fill=cIn},
  yc/.style ={cell, fill=cOut},
  tc/.style ={cell, fill=cOut2},
  qc/.style ={cell, fill=cQ, text=cQD, font=\small\bfseries},
  fr/.style ={line width=0.9pt, rounded corners=0.8pt},
  rl/.style ={font=\small, anchor=east},
  hd/.style ={font=\normalsize\bfseries, anchor=south},
  cn/.style ={font=\small\itshape, anchor=south},
  bx/.style ={line width=0.8pt, rounded corners=1.6pt,
              font=\scriptsize, align=center, inner sep=2.6pt},
  nb/.style ={bx, draw=cNetD, fill=cNet, text=black},
  qb/.style ={bx, draw=cQD,   fill=cQ,   text=cQD!80!black},
  hz/.style ={bx, draw=cInD,  fill=cIn,  text=black, minimum width=32mm},
  hy/.style ={bx, draw=cOutD, fill=cOut, text=black, minimum width=36mm},
  ar/.style ={-{Latex[length=1.7mm]}, line width=0.85pt}
]
\def\ZA{0}\def\ZB{1}\def\ZC{2}
\def\VA{3.4}\def\VB{4.4}\def\VC{5.4}
\def\TA{6.4}\def\TB{7.4}\def\TC{8.4}

\node[nb, font=\footnotesize, minimum width=55mm, inner ysep=1.4pt] (net) at (3.45,5.35)
     {\textbf{Network}\ \ \scriptsize Case14, Case30, Case33bw, Case70, Case118};
\node[hz] (hz) at (1.00,3.62) {\textbf{Historical measurements}};
\node[hy] (hy) at (5.90,3.62) {\textbf{Historical estimates}, $v_i$, $\Delta\theta_{ij}$};
\draw[ar, cInD]  (hz.north |- net.south) -- (hz.north);
\draw[ar, cOutD] (hy.north |- net.south) -- (hy.north);
\draw[ar, -{Latex[length=1.2mm]}, cInD]  (hz.south) -- (1.00,2.80);
\draw[ar, -{Latex[length=1.2mm]}, cOutD] (hy.south) -- (5.90,2.80);

\begin{scope}[yshift=-1.1mm]
\node[cn] at (0.50,2.00) {Readings, Model input, $\mathbf{X}_{\mathrm{ctx}}$};
\node[cn] at (5.90,2.00) {States, the answers, $\mathbf{Y}_{\mathrm{ctx}}$};
\draw[decorate, decoration={brace, amplitude=4pt}, cInD!80,  line width=0.7pt]
      (-0.48,1.58) -- (2.48,1.58);
\draw[decorate, decoration={brace, amplitude=4pt}, cOutD!80, line width=0.7pt]
      ( 2.92,1.58) -- (8.88,1.58);

\foreach \x/\t in {\ZA/$z_1$, \ZB/$\cdots$, \ZC/$z_m$}
  \node[font=\normalsize, text=black] at (\x,1.00) {\t};
\foreach \x/\t in {\VA/$v_1$, \VB/$\cdots$, \VC/$v_n$,
                   \TA/$\Delta\theta_1$, \TB/$\cdots$, \TC/$\Delta\theta_L$}
  \node[font=\normalsize, text=black] at (\x,1.00) {\t};

\foreach \y/\lab/\hh/\ff/\a/\b/\c/\e/\f/\g/\p/\q/\r in {%
  0/snapshot 1/3.3mm/\footnotesize/1.01/$\cdots$/-0.18/1.000/$\cdots$/0.981/0.031/$\cdots$/0.019,
  -0.85//2.6mm/\tiny/\rotatebox{90}{$\cdots$}/\rotatebox{-45}{$\cdots$}/\rotatebox{90}{$\cdots$}/\rotatebox{90}{$\cdots$}/\rotatebox{-45}{$\cdots$}/\rotatebox{90}{$\cdots$}/\rotatebox{90}{$\cdots$}/\rotatebox{-45}{$\cdots$}/\rotatebox{90}{$\cdots$},
  -1.70/snapshot $N$/3.3mm/\footnotesize/1.02/$\cdots$/-0.19/1.001/$\cdots$/0.984/0.029/$\cdots$/0.017}{
  \node[rl] at (-0.56,\y) {\lab};
  \node[zc, minimum height=\hh, font=\ff] at (\ZA,\y) {\a}; \node[zc, minimum height=\hh, font=\ff] at (\ZB,\y) {\b}; \node[zc, minimum height=\hh, font=\ff] at (\ZC,\y) {\c};
  \node[yc, minimum height=\hh, font=\ff] at (\VA,\y) {\e}; \node[yc, minimum height=\hh, font=\ff] at (\VB,\y) {\f}; \node[yc, minimum height=\hh, font=\ff] at (\VC,\y) {\g};
  \node[tc, minimum height=\hh, font=\ff] at (\TA,\y) {\p}; \node[tc, minimum height=\hh, font=\ff] at (\TB,\y) {\q}; \node[tc, minimum height=\hh, font=\ff] at (\TC,\y) {\r};
}
\draw[fr, cInD!65]  (-0.49,0.56) rectangle (2.49,-2.17);
\draw[fr, cOutD!65] ( 2.91,0.56) rectangle (8.89,-2.17);
\draw[cOutD!45, line width=0.6pt] (5.90,0.56) -- (5.90,-2.17);

\node[rl] at (-0.56,-3.20) {query};
\node[zc] at (\ZA,-3.20) {1.00}; \node[zc] at (\ZB,-3.20) {$\cdots$};
\node[zc] at (\ZC,-3.20) {-0.20};
\foreach \x in {\VA,\VB,\VC,\TA,\TB,\TC} \node[qc] at (\x,-3.20) {?};
\draw[fr, cInD!65] (-0.49,-2.73) rectangle (2.49,-3.67);
\draw[black!85, line width=1.0pt, dash pattern=on 2.4pt off 1.8pt]
      (-0.49,-2.36) -- (2.49,-2.36);
\draw[black!85, line width=1.0pt, dash pattern=on 2.4pt off 1.8pt]
      ( 2.91,-2.36) -- (8.89,-2.36);

\node[draw=cQD, line width=1.0pt, rounded corners=1.2pt, fill=none,
      minimum width=45.0mm, minimum height=4.6mm, inner sep=0pt,
      drop shadow={shadow xshift=0.45mm, shadow yshift=-0.45mm,
                   opacity=0.30, fill=cQD}] (svbox) at (5.90,-3.20) {};

\node[qb, font=\small] (qsrc) at (0.60,-5.00) {\textbf{QUERY}};
\node[cn, anchor=north, font=\normalsize\itshape, text=cQD!70!black]
      at (0.10,-5.62) {New Meter Readings};
\draw[ar, cQD] (qsrc.north) -- (0.60,-3.75);
\node[draw=cQD, line width=0.8pt, rounded corners=1.6pt, fill=cQ!45,
      inner xsep=3pt, inner ysep=1.4pt, font=\small, text=black] (sv) at (5.90,-5.00)
     {$\hat{\mathbf{x}}_q=[\,\hat v_1\,\cdots\,\hat v_n\;\Delta\hat\theta_1\,\cdots\,
       \Delta\hat\theta_L\,]^{\top}$};
\node[cn, anchor=north, font=\normalsize\itshape, text=cQD!70!black]
      at (5.90,-5.62) {Predicted State Vector};
\draw[ar, cQD, line width=0.8pt] (svbox.south) -- (5.90,-4.52);
\end{scope}
\end{tikzpicture}
\vspace{-2.2em}
\caption{Proposed in-context state estimation via TFMs. The query consists of the current meter readings for which the state vector is to be predicted.}
\label{fig:method}
\vspace{-1.8em}
\end{figure}

The proposed ICTP instead poses SE as a tabular prediction task using two pretrained TFMs, TabPFN\cite{tabpfn} and TabICL\cite{tabicl}. The context table holds $N$ historical snapshots of the network being estimated, one per row (Fig.~\ref{fig:method}). Measurements form the input columns $\mathbf{X}_{\mathrm{ctx}}$. These are bus voltage magnitudes, active and reactive bus injections, and active and reactive branch flows at the from-end, all carrying noise. No phase angle is measured. The targets $\mathbf{Y}_{\mathrm{ctx}}$ are corresponding state estimates, one column per bus for the voltage magnitude and one per line for the angle difference $\Delta\theta_{ij}$. Each context row is a past snapshot in which all meters were in service, and its targets are the WLS estimates obtained at that time. All errors are computed against the true power-flow states. Predicting angle differences rather than absolute angles lowers the angle error by 1.1 to 1.9$\times$ at full observability, on all networks. The query is the current measurement vector $\mathbf{z}_q$, and each of the $C$ target columns is predicted in its own forward pass:
\begin{equation}
\label{eq:icl}
\hat{y}_{q,k} \;=\; f_\theta\!\left(\mathbf{z}_q \,;\, \{(\mathbf{z}^{(t)}, y_{t,k})\}_{t=1}^{N}\right),
\qquad k = 1,\dots,C,
\end{equation}
where $\mathbf{z}^{(t)}$ is row $t$ of $\mathbf{X}_{\mathrm{ctx}}$, $y_{t,k}$ is the $k$th entry of row $t$ of $\mathbf{Y}_{\mathrm{ctx}}$, and $f_\theta$ is the TFM with its pretrained parameters $\theta$ held fixed.
Equation~\eqref{eq:icl} uses no $\mathbf{H}$, admittance matrix, or network connectivity, and the same weights serve any network. Connectivity enters only afterward, when absolute angles are recovered from the predicted differences by a linear least-squares solve on the branch-bus incidence matrix with the slack angle fixed.

\vspace{-1.05em} 
\section{Results and Discussions}
\label{sec:results}
We evaluate on two fronts. First, on the 30- and 118-bus transmission systems and the 33-bus radial distribution system, we compare ICTP against WLS and its robust variants \cite{pandapower}.
Second, we compare ICTP with DSS$^2$ \cite{habib2024dss2} and TAPINE \cite{tapine2026} under the conditions their authors report.

For the first front, each snapshot is one operating point. Real loads are drawn from a Gaussian with 30\% standard deviation about nominal, each load's reactive-to-active ratio from the case file perturbed by a Gaussian with 20\% standard deviation, and power flow is solved before adding noise. Meters measure branch flows, bus injections, and bus voltage magnitudes, all present at full metering. \textbf{Base noise} is zero-mean Gaussian with standard deviation $\sigma$. Voltage magnitudes take an absolute $\sigma$ of 0.002~p.u.\ while injections and flows are perturbed by 2\% of their true value, and $\sigma$ is 2\% of the resulting noisy reading. A floor on $\sigma$ keeps the weight finite at zero-injection buses.
WLS receives $\sigma$ for each measurement, while the ICTP receives only the noisy measurements. The context holds $N$ past snapshots with their corresponding state estimates. ICTP performs 100 estimations from a single context table, and errors are the mean absolute error (MAE) over all buses and those 100, over converged snapshots only. 
The context mean, each bus's average over its context rows, is the baseline a useful estimator must beat. The tests are:
\vspace{-0.25em} 
\begin{itemize}
    \item \textbf{Test A, meter loss at base noise.} The context targets stay as they are, and measurements are removed from the context and query input columns: \textbf{A1)} a percentage of a measurement type randomly across buses,
    \textbf{A2)} dark buses, every measurement at the $k$ highest-degree buses, and the flows on their incident branches, for $k=1,2,3,5,8$, never the slack, and \textbf{A3)} is A1 without restriction of measurement types.
    \item \textbf{Test B, noise variations.} \textbf{B1)} At full metering, $\sigma$ is scaled by 5 and 2.5 for voltage magnitude and power measurements respectively, and WLS receives the scaled $\sigma$. \textbf{B2)} Laplace noise at same $\sigma$, scale $\sigma/\sqrt{2}$.
    \item \textbf{Test C, bad data.} Every measurement at 3 or 6 random buses, and the flows on their incident lines, is biased by $3\sigma$ or $8\sigma$ with an independent random sign, while the context stays clean. WLS here runs with bad-data detection ($\chi^2$ and largest normalized residual). Tests A and B use plain WLS.
\end{itemize}
Context size is the one design choice. At full metering, the error falls 2.4 to 22$\times$ from 10 to 100 rows, and at most 1.7$\times$ from 100 to 500, so $N=500$ is used throughout.

\begin{figure}[t]
\centering
\vspace{-2.5em} 
\includegraphics[width=\columnwidth]{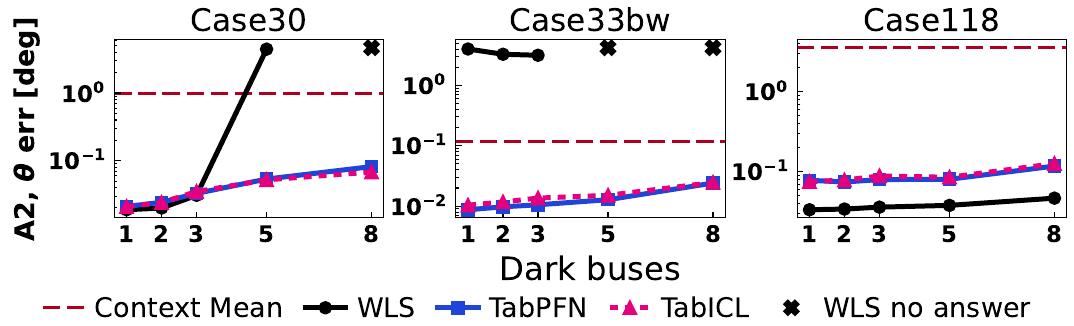}\\[-1pt]
\includegraphics[width=\columnwidth]{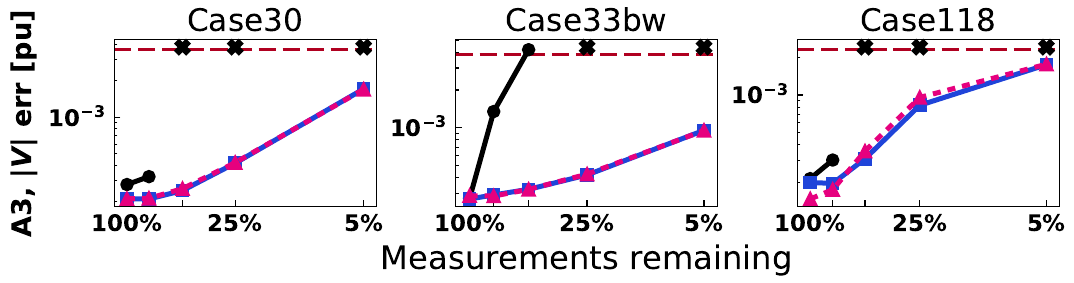}
\vspace{-2.3em}
\caption{Test A results. \textbf{Top}, A2 angle error. \textbf{Bottom}, A3 voltage error. Crosses mark levels where WLS returned no estimate, the red line is the context mean.}
\label{fig:testa}
\vspace{-1.8em}
\end{figure}

First, \textbf{Test A1} showed that any single measurement type can be thinned heavily on every network at a cost in accuracy, not convergence, since $\mathbf{H}$ stays at full rank at every level. Losing flow meters down to 5\% of the total costs 118-bus the most, raising the WLS angle error 7.7$\times$ to 0.249$^\circ$. Losing injection meters to the same 5\% raises it only 1.6$\times$ to 0.051$^\circ$, and losing voltage meters down to 10\% moves it by at most 5\%. ICTP is far less sensitive, varying by at most 0.006$^\circ$ across the injection sweep.

In \textbf{Test A2}, as in Fig.~\ref{fig:testa}(top), the 33-bus is the worst hit because a single dark bus leaves the radial network unobservable, and the WLS angle error exceeds the context-mean error.
The ICTP error rises gradually on all networks and never exceeds 4.4$\times$ its full-metering angle error, or 1.3$\times$ for voltage.
\textbf{Test A3} sweep is shown in Fig.~\ref{fig:testa}(bottom), in which WLS on 30-bus and 118-bus converged on all 100 snapshots from 100\% to 75\% and returned no estimate with $\leq 50\%$ measurement available. ICTP returned a full state vector at every level, and stayed below the context mean in Fig.~\ref{fig:testa} by 11 to 57$\times$ at full metering and 1.3 to 4.0$\times$ at 5\% metering.

With all measurements, ICTP's voltage error is 0.69 to 1.05$\times$ that of WLS and its angle error 1.46 to 2.03$\times$, with both ratios moving in ICTP's favor as measurements reduce. ICTP sees context, while WLS does not, so for a fair comparison we also fit ridge regression on the same context. At 500 rows ridge is competitive, within 1.6$\times$ of ICTP on voltage across the three
networks, and lower on angle on the two smaller feeders. At 50 rows it is not, and TabPFN
is 4.3$\times$ better on voltage and 7.3$\times$ better on angle on the 118-bus system,
since it is pretrained whereas ridge refits for every network at each metering level.
At every test level where $\mathbf{H}$ is full rank, WLS converged on all 100 snapshots
and $\mathrm{cond}(\mathbf{G})$ stayed below $10^{13}$. Wherever the rank fell short,
it exceeded $10^{32}$, and WLS either returned nothing or answered on some snapshots with an error far above the context mean. Having $m>n$ therefore guarantees neither a good estimate nor any estimate at all.
\begin{figure}[t]
\centering
\vspace{-1.9em}
\includegraphics[width=\columnwidth]{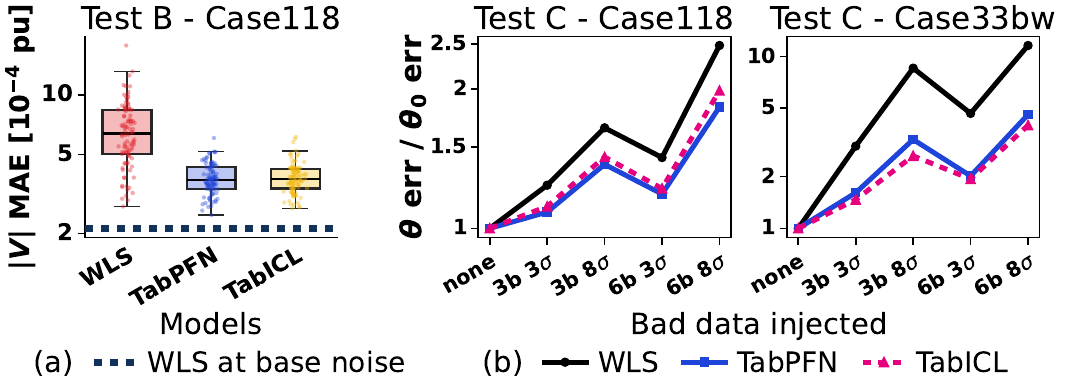}
\vspace{-2em}
\caption{\textbf{Left} Test B1, voltage error on 118-bus at raised noise, dotted line, WLS (base noise). \textbf{Center and right}, Test C angle error ratio to clean value.}
\label{fig:bd}
\vspace{-1.35em}
\end{figure}

 \textbf{Test B1}, in Fig.~\ref{fig:bd} (left), raises the noise at full metering. Every method's error increases by a similar factor against its own base-noise result, 1.8 to 3.1$\times$ in voltage and 2.0 to 2.7$\times$ in angle, with neither WLS nor ICTP consistently worse. For \textbf{Test B2}, we report that Laplace noise at the base $\sigma$ leaves each method at 0.86 to 1.15$\times$ its base Gaussian noise error. In \textbf{Test C}, WLS angle estimates degrade more than ICTP at every level on every network (Fig.~\ref{fig:bd} (right)). With six buses having 8$\sigma$ bias (bad data), the WLS angle error reaches 2.5 to 12.7$\times$ of unbiased situation error, against 1.8 to 5$\times$ for ICTP. The voltage ratios are 2.5 to 5.2$\times$ for WLS and 1.3 to 3.5$\times$ for ICTP. For the same test, among the other WLS variants in \textit{pandapower 3.4.0}~\cite{pandapower}, IRWLS,
Schweppe-Huber GM, quadratic-constant, and Huber answer at most 1 of 100 snapshots. Least-absolute-value answers 97 to 100 and beats ICTP at one of the four biased levels, losing at the other three.

\begin{figure}[t]
\centering
\includegraphics[width=\columnwidth]{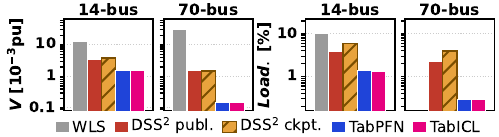}
\vspace{-2.em}
\caption{TU-Delft DSS$^2$ comparison on their 1080-snapshot test file with their published checkpoint. Voltage RMSE in $10^{-3}$\,p.u. and line loading in \%. WLS's 70-bus line loading, $1.3\times10^{4}$\,\%, is off-scale and omitted.}
\label{fig:tudelft}
\vspace{-1.8em}
\end{figure}

Now for the second testing front. For the TU-Delft model \cite{habib2024dss2} comparison, we ran the published DSS$^2$ checkpoint and WLS on their 1080-snapshot test file for the CIGRE 14-bus and Oberrhein 70-bus distribution networks.
TabPFN and TabICL receive 500 context rows from TU-Delft's training samples and predict out of sample without training. In the context, true states are used instead of past estimates, since WLS converged on only 227 of the 500 context snapshots.
Both have lower voltage error than DSS$^2$ and WLS on both feeders in Fig.~\ref{fig:tudelft},
while WLS converged only for 501 of 1080 snapshots on the 70-bus feeder whose $\mathbf{H}$
is full rank, so the failure is numerical rather than a lack of measurements, and its bar
averages those 501.
\begin{figure}[!t]
\centering
\vspace{-1.5em}
\includegraphics[width=\columnwidth]{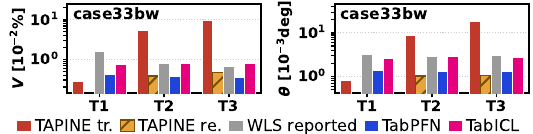}
\caption{Feeder reconfiguration under TAPINE's convention \cite{tapine2026}. WLS values are as cited, tr. refers to TAPINE's transferred model, and re.\ retrained.}
\vspace{-1.6em}
\label{fig:tapine}
\end{figure}
The second learned estimator is TAPINE~\cite{tapine2026}, which tests the system on reconfigured topologies, where sets T1, T2, and T3 contain 11, 20, and 25 feeder topologies built from one, two, and three simultaneous switch pairs (as in their Section~IV-A). Fig.~\ref{fig:tapine} plots the TAPINE \textit{transferred model} and TAPINE \textit{retrained model} on each set. Their T1-trained model degrades 19 and 35$\times$ in voltage on T2 and T3, and recovers only after retraining on each set. For ICTP, context and query snapshots are disjoint draws from each topology, 500 WLS-estimated context labels per topology.
With TabPFN, voltage MAPE stays within 0.358 to 0.421 $[10^{-2}\,\%]$, TAPINE's own metric and unit, across the three sets, and is below its retrained model at both T2 and T3.
ICTP therefore matches a learned estimator that must be retrained for each topology set, needing only labeled context from that topology and no weight update.

\vspace{-0.8em} 
\section{Conclusion}
We posed state estimation as an in-context tabular prediction problem, where past measurements and their state estimates form the context, and a new measurement vector is the query. The same pretrained weights estimate the state of five networks. Across the three stress tests, WLS converged wherever $\mathbf{H}$ was full rank, and when rank fell short, it returned either nothing or an estimate whose error exceeded the context mean. Neither the measurement count nor the WLS bad-data checks reveal the rank loss, so $m>n$ doesn't guarantee a good estimate, or any estimate at all.
ICTP returned a full state vector at every level and stayed below the context mean throughout. It also degrades less than WLS with bad-data detection under injected bias, and reaches or beats two published learned estimators on their own benchmarks without retraining. ICTP costs 239 to 393\,s for 100 estimations on the 118-bus system against 81\,s for WLS, so reducing the per-target loop is left to future work.

\vspace{-0.8em} 
\bibliographystyle{IEEEtran}  
\bibliography{main} 
\end{document}